\documentstyle[11pt,paspconf,epsf]{article}
\def\refer#1{#1}

\begin{document}

\title{Optical and Near-IR Field Luminosity Functions}
\author{Jon Loveday}
\affil{Astronomy \& Astrophysics Department, University of Chicago, 
5640 S Ellis Ave, Chicago, IL 60637, USA}

\begin{abstract}
We present preliminary measurements of the $b_J$ and $K$-band luminosity 
functions (LFs) of field galaxies obtained from optical and $K$-band imaging 
of a sample of galaxies selected from the Stromlo-APM Redshift Survey.
The $b_J$ LF is consistent with that previously published from
photographic data.
The $K$-band LF has been estimated over a range of 12 magnitudes and is
reasonably well fit by a Schechter function with faint-end slope 
$\alpha = -1.2$.
\end{abstract}

\keywords{galaxies: distances and redshifts
--- galaxies: luminosity function, mass function
--- cosmology: observations
--- surveys}

\section{Introduction}

Deep, near-infrared $K$-band ($2.2\mu$m) galaxy surveys are a powerful tool for
studying galaxy evolution
(eg. Gardner et al.\ 1993, Cowie et al.\ 1994, Glazebrook et al.\ 1995).
Compared to blue-optical light, near-infrared light is a better tracer
of mass in evolved stars and the correction for redshift
dimming (the ``$k$-correction'') is approximately independent of
morphological type.
The rapid evolution in galaxy luminosity apparent in the $b_J$ band is not
seen in the $K$ band.
However, it is vital to have a reliable determination of the $K$-band
luminosity function for {\em nearby} galaxies in order to interpret
faint galaxy counts and to calculate the clustering of $K$-selected galaxy
samples.
The largest local $K$-band sample of galaxies with redshifts
is that of Gardner et al.\ (1997).
They measured redshifts for 510 galaxies selected from a $K$-band limited
survey covering 4.4 square degrees.
Since their survey is flux-limited, the majority of galaxies have
$K$-band luminosities close to $L_K^*$.
Thus they are able to measure the $K$-band luminosity function over a range
of only 5 magnitudes, and the faint-end slope of their best-fit Schechter
function, so important for predicting galaxy number counts,
is poorly constrained (their Figure~1).

One can improve on current estimates of the $K$-band LF without a huge
investment of telescope time by observing galaxies selected by their
intrinsic luminosity rather than their apparent flux.
The Stromlo-APM galaxy survey (Loveday et al.\ 1996) is an ideal source
for a new determination of the joint optical/near-infrared luminosity function
$\phi(L_{b_J}, L_K)$ since redshifts have already
been measured for 1797 galaxies with $b_J < 17.15$
over a very large volume of space.
The solid angle of the survey is 1.3 sr and the median redshift is
about 15,300 km/s.
One can make use of the fact that $K$ and $b_J$ luminosities are correlated, 
so that we can preferentially select galaxies of high and low luminosity and
thus sample the luminosity range more evenly than a flux-limited sample.
We are thus able to measure the luminosity function to fainter
luminosities than from a flux-limited sample of similar size.

\section{Sample Selection} \label{sec:sample}

Our aim in selecting a subset of Stromlo-APM galaxies for which to obtain
$K$-band photometry was to sample the magnitude range 
$-22 \le M_{b_J}\footnote{Throughout, we assume a Hubble constant of
$H_0 = 100$ km/s/Mpc.} \le -13$
as uniformly as possible.
An added complication in defining the sample arose because we wished
to obtain optical CCD images for the same sample of galaxies.
One planned use of this optical imaging is to measure morphological
parameters for a representative sample of galaxies at low redshift 
in order to compare with HST observations of galaxies
at high redshift ($z > 0.4$).
To obtain comparable linear resolution to the HST data required observing
galaxies at $z < 0.04$, assuming ground-based seeing of 1.3 arcsecond.
Our ``primary'' sample thus consists of galaxies at redshifts $z < 0.04$.
We divided the magnitude interval $-22 \le M_{b_J} \le -13$ into 90 bins
each of width 0.1 mag.
We then randomly selected up to six galaxies from the Stromlo-APM survey with
$z < 0.04$ in each bin.
Due to its redshift limit of $z < 0.04$, this primary sample
contains rather few galaxies brighter than $M_{b_J} = -20$.
We therefore formed a supplementary sample, consisting of galaxies at
$z > 0.04$ to ``top up'' each magnitude bin, where possible,
to six galaxies.
This supplementary sample consists entirely of galaxies with $M_{b_J} < -20$.
The primary sample contains 283 galaxies, and the supplementary sample
contains 80 galaxies, giving a total sample size of 363 galaxies.

\section{Observations}

$K$-band imaging of the above sample of galaxies was carried out at the
Cerro Tololo Interamerican Observatory (CTIO) 1.5m telescope using
the CIRIM infrared array over the nine nights 1996 August 31 -- 
September 4 and 1997 October 19--22.
The pixel size at $f/7.5$ is $1.16''$, allowing most galaxies
to be observed at 9 non-overlapping positions on the $256 \times 256$ array.
Total integration time for each galaxy was 300s.
The infrafred frames were reduced using {\sc IRAF}, and 
photometry was performed using SExtractor 2.0.8 (Bertin \& Arnouts 1996)
with the ``mag\_best'' option.
Magnitude errors were estimated by combining in quadrature SExtractor's
estimate of the error from photon statistics and the difference between
magnitudes measured using local and global estimates of the sky background.
343 galaxies were observed with an estimated $K$-band magnitude error of
less than 0.3 mag (rms mag error = 0.06 mag)
and were calibrated using standard stars observed
from the list of Elias et al.\ (1982).

\begin{figure}
\plotone{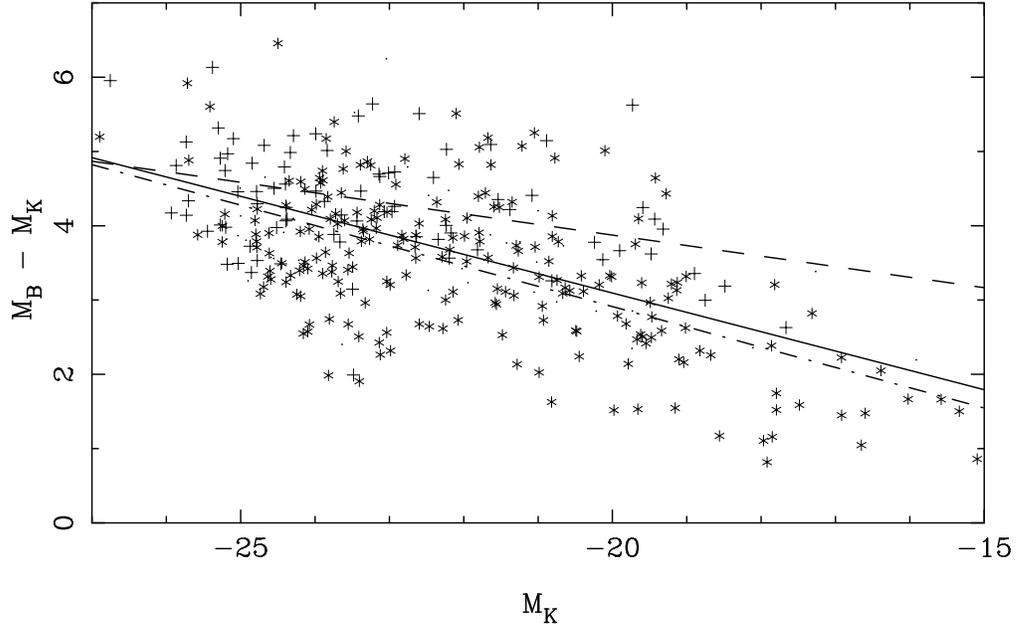}
\caption{$b_J - k$ versus $M_k$ colour-magnitude plot using
	APM $b_J$ magnitudes.
	Plus signs represent early-type galaxies, asterisks late-type
	galaxies and dots represent unclassified galaxies.
	The solid line shows a least-squares fit to all galaxies, 
	the dahsed line a fit to early types and the dot-dashed line a fit
	to late type galaxies.}
\label{fig:colmag}
\end{figure}

Figure~\ref{fig:colmag} shows the rest-frame $(b_J - k)$ versus $M_K$
colour-magnitude relation for our data.
The fit to all galaxies is given by
\begin{equation}
(b_J - k) = -0.260 \times M_K - 2.11,\ \sigma = 0.86. \label{eqn:colmag}
\end{equation}
We assume a $K$-band k-correction of $-2.5 z$ for all galaxy types.

\begin{figure}[htbp]
\parbox{7cm}{
\epsfxsize=7cm
\epsfbox{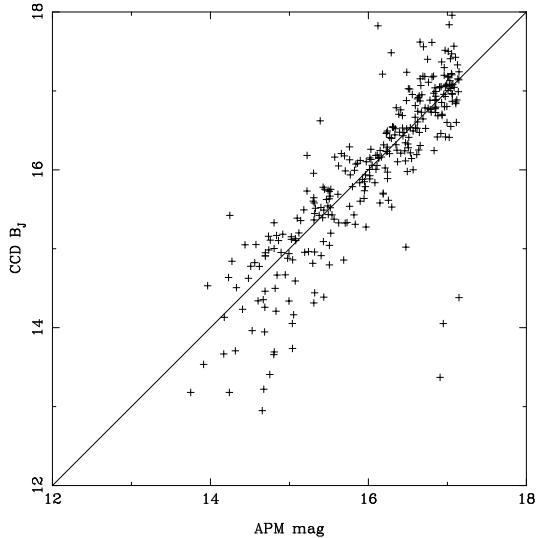}
}
\parbox{6cm}{
\caption{CCD $b_J$ magnitudes plotted against APM magnitudes.}
\label{fig:bjmags}
}
\end{figure}

Optical imaging was performed in the $U$, $B$ and $R$ bands using the CTIO 
1.5m telescope with a Tex $2048^2$ CCD over the ten nights 1996 September 7--16.
Integration times were 120s in $R$, 240s in $B$ and 120s in $U$.
The $U$-band exposures were too short to provide accurate galaxy photometry
but were taken under photometric conditions, allowing the possibility of
later calibration of deeper, non-photometric $U$-band observations.
Galaxy photometry was done using an earlier beta-release (1.2b9b) of SExtractor 
and so the photometry presented here is only preliminary.
300 galaxies have reliable $B$ and $R$ magnitudes, and are calibrated with 
Landolt (1992) standards.
The colour equations of Couch \& Newell (1980) were used to obtain a $b_J$
magnitude from $B$ and $R$.
Figure~\ref{fig:bjmags} plots these CCD $b_J$ magnitudes against APM $b_J$ 
magnitudes.
The mean and rms APM $-$ CCD magnitude is $\Delta m = 0.09 \pm 1.05$.
This scatter is larger than the 0.3 mag scatter for the full, $m > 15$
Stromlo-APM
sample (Loveday et al.\ 1992) since 1) galaxies with APM magnitude brighter
than 15 are included in the current sample, these are badly saturated on the
photographic plates, and 2) the preferential sampling of galaxies of very
high and low luminosity means that galaxies with poor APM magnitudes are
more likely to be included.
For example, the outlying galaxies in the lower-right of 
Figure~\ref{fig:bjmags} are due to the APM machine measuring just a small
part of a large spiral galaxy, and thus grossly underestimating the galaxy's
luminosity.

\section{Luminosity Functions}
\label{sec:method}

When one has a sample selected on one quantity $m_1$ (in our case 
APM $b_J$ magnitude) and wishes to estimate the luminosity 
function for another quantity $m_2$ (eg. CCD $b_J$ or $K$ magnitude),
the best way to proceed is to calculate a bivariate luminosity function 
(BLF) $\phi(L_1,L_2)$ allowing for any selection effects in $L_1$
and then to integrate over $L_1$ to obtain $\phi(L_2)$.
One can estimate the shape of $\phi(L_1,L_2)$,
independently of inhomogeneities in the galaxy distribution using
the maximum likelihood method of \refer{Sandage, Tamman and Yahil (1979)}.
The probability of seeing a galaxy with luminosities $L_1^i$ and $L_2^i$
at redshift $z_i$ is given by
\begin{equation}
p_i = \phi(L_1^i, L_2^i) S(L_1^i) \left/ 
      \int_{{L_2}_{\rm min}(z_i)}^{{L_2}_{\rm max}(z_i)}
      \int_{{L_1}_{\rm min}(z_i)}^{{L_1}_{\rm max}(z_i)}
      \phi(L_1, L_2) S(L_1) dL_1 dL_2.\right.
\end{equation}
The function $S(L_1)$ accounts for the known selection in $L_1$ 
and the luminosity limits ${L_1}_{\rm min}(z_i)$ and 
${L_1}_{\rm max}(z_i)$
are the minimum and maximum luminosities observable at
redshift $z_i$ in a sample limited by apparent $m_1$ magnitude.
If there are no flux limits in the $m_2$-band, then
the integral over $L_2$ runs from $0$ to $+\infty$.
The maximum-likelihood shape of the BLF $\phi(L_1,L_2)$ is estimated by 
maximizing
the likelihood ${\cal L} = \prod_{i=1}^{N_g} p_i$ (the product of the 
individual probabilities $p_i$ for the $N_g$ galaxies in the sample) 
with respect to the parameters describing the BLF.

In practice, we do not have a good {\em a priori} parametric model for
$\phi(L_1,L_2)$, and so instead we measure $\phi(L_1,L_2)$
in a non-parametric way using an extension of the 
\refer{Efstathiou, Ellis and Peterson (1988)} stepwise maximum likelihood
(SWML) method.
\refer{Sodr\'e and Lahav (1993)} have extended the SWML method to estimate
the bivariate diameter-luminosity function and to allow for sample 
incompleteness.
We adopt their extension of the SWML estimator here, including the sampling
function $S(L_1)$ separately for the primary and supplementary galaxy samples.
We normalise our LFs to the mean density of galaxies with 
$-22 \le M_{b_J} \le -13$ in the full Stromlo-APM sample, 
$\bar{n} = 0.071 h^3{\rm Mpc}^{-3}$, calculated as described by Loveday et al.
(1992).

Once one has obtained the SWML estimate of $\phi(L_1,L_2)$, one can 
integrate over $L_1$ to obtain $\phi(L_2)$ and then
fit a given functional form, eg. a Schechter (1976) function,
by least-squares.

\subsection{Test of the Method}

We have tested the above procedure by using it to
estimate the $K$-band luminosity function from a set of Monte Carlo 
simulations.
We generated nine mock Stromlo surveys by a \refer{Soneira and Peebles (1978)}
hierarchical clustering simulation.
Each galaxy in the simulation was assigned a $K$-band luminosity 
drawn at random from
a Schechter function with $\alpha = -1.21$ and $M_K^* = -24.7$.
Each galaxy was then assigned a $b_J$ magnitude according to our observed
colour-luminosity relation (\ref{eqn:colmag}).
Galaxies were selected on their apparent $b_J$ magnitude, $b_J < 17.15$.
This process was repeated until each simulation contained 2000 galaxies.
We then sampled each simulation by absolute $M_{b_J}$ magnitude as described
in \S\ref{sec:sample}, finally yielding an average of 359 galaxies
per simulation.
We calculated the K-band luminosity function $\phi(L_K)$ for each
simulation as described in \S\ref{sec:method} and fit a Schechter function
to each by least squares.
Averaging over the nine simulations, and estimating the BLF in bins of width
0.5 mag, we measure mean and rms Schechter
function parameters $\alpha = -1.16 \pm 0.06$, $M^* = -24.4 \pm 0.6$.
The errors on the mean values are $\sqrt 9$ times smaller than the quoted
rms scatter between the simulations.
Thus our estimate of $M^*$ is biased $1.5\sigma$ too faint and
$\alpha$ is overestimated (too shallow) by about $2.5\sigma$.
However, our estimates are within the $1\sigma$ error from a single 
realisation.

\begin{figure}
\plotone{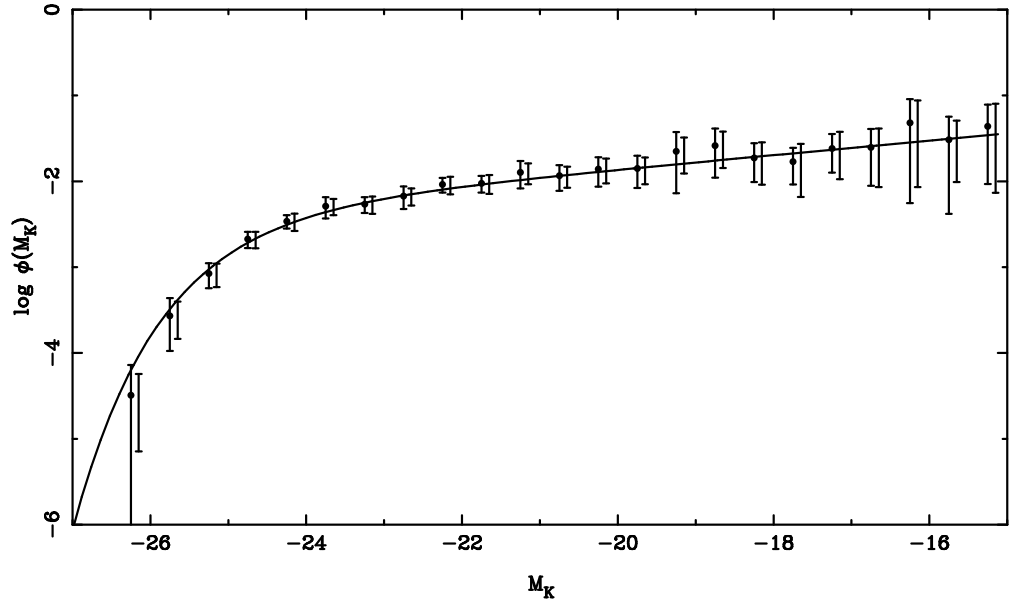}
\caption{The $K$-band luminosity function estimated from nine 
Soneira-Peebles simulations (points) along with the input LF (curve).
\label{fig:test}}
\end{figure}

The SWML estimates of the $K$-band LF for the simulations
are shown in Figure~\ref{fig:test}.
The points show the mean SWML estimate of $\phi(L_K)$ from
the nine simulations and the curve shows the input Schechter function with
shape $\alpha = -1.21$, $M^*_K = -24.7$.
The error bars going through the data points show the rms scatter between 
realisations.
The error bars offset slightly to the right show the mean error predicted by
the covariance matrix (see Efstathiou et al) and are in reasonable agreement 
with the rms scatter between realisations.

Overall, we find that our procedure for estimating $\phi(L_K)$ from a
sample limited by apparent $b$ magnitude and futher selected by absolute
$B$ magnitude provides a robust and only weakly biased estimate of the 
$K$-band LF over a wide range of absolute magnitudes.

\subsection{$\phi(L_{b_J})$}

\begin{figure}
\plotone{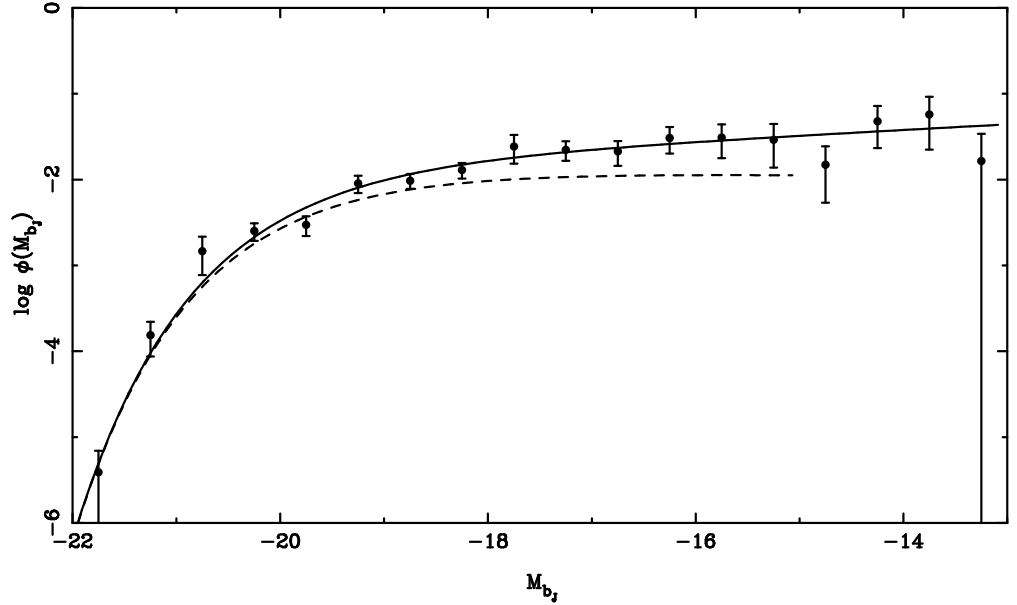}
\caption{The $b_J$ luminosity function estimated from our CCD magnitudes
(filled symbols).
Also shown (dashed line) is the Schechter function fit by Loveday et al.\ (1992)
from photographic magnitudes.}
\label{fig:bjlf}
\end{figure}

Our estimated $b_J$-band luminosity function is shown in 
Figure~\ref{fig:bjlf}.
The curve shows a Schechter function fit to the SWML estimate using least
squares and allowing for finite bin width.
The best-fit Schechter parameters are $\alpha = -1.16$, $M_{b_J}^* = -19.52$ 
and $\phi^* = 0.018 h^3 {\rm Mpc}^{-3}$.
Also shown in this figure is the Schechter function fit by Loveday et al.\ (1992)
from photographic magnitudes.
Despite the large scatter in our CCD versus APM $b_J$ magnitudes
(Fig.~\ref{fig:bjmags}), we find that the two estimates of the shape of
$\phi(M_{b_J})$ are in reasonable agreement.
Although the new estimate has slightly steeper faint-end slope, the slopes
are in fact consistent within the $1\sigma$ uncertainties.

\subsection{$\phi(L_K)$}

\begin{figure}
\plotone{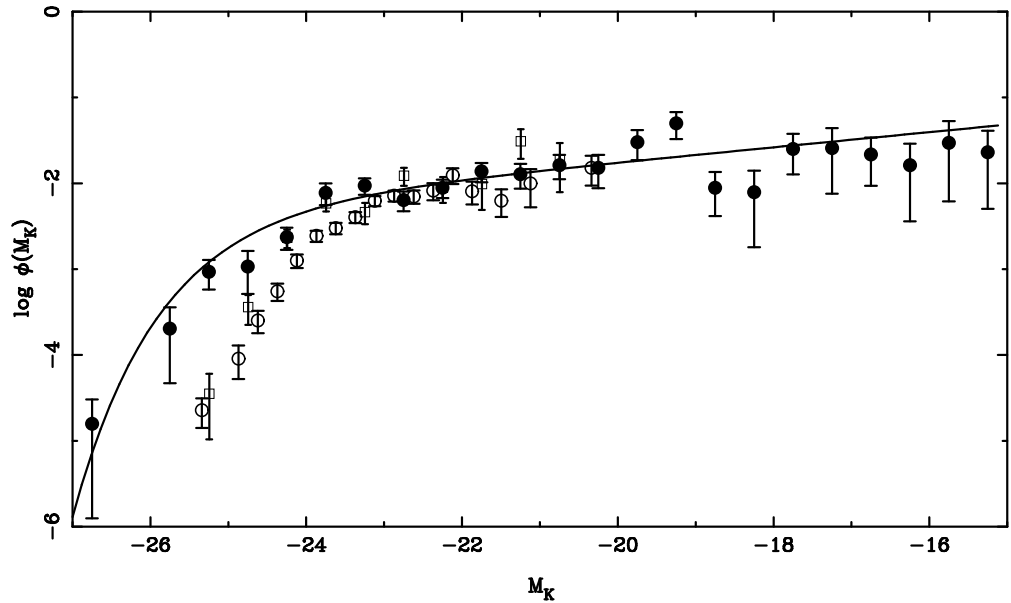}
\caption{The $K$-band luminosity function estimated from our sample
(filled symbols).
Also shown are the results of Gardner et al.\ (open circles) and Szokoly et al.
(open squares).}
\label{fig:klf}
\end{figure}

Our estimated $K$-band luminosity function is shown in 
Figure~\ref{fig:klf}.
The curve shows a Schechter function fit to the SWML estimate using least
squares and allowing for finite bin width.
The best-fit Schechter parameters are $\alpha = -1.22$, $M_K^* = -24.73$ and
$\phi^* = 0.0073 h^3 {\rm Mpc}^{-3}$.
Also shown in this figure are recent estimates of the $K$-band LF from the
$K$-selected samples of Gardner et al.\ (1997) and of Szokoly et al.\ (1998).
These estimates are in good agreement with ours at $M_K \approx -22$
but fall below our new estimate at the bright end.
The reason for this is almost certainly the fact that our sample is selected
in $b_J$, whereas the Gardner et al.\ and Szokoly et al.\ samples are 
$K$-selected.
In a blue-selected sample, the bright end of the $K$-band LF will be dominated
by red galaxies and the faint end dominated by blue galaxies.
Since red galaxies tend to be luminous ellipticals, this would explain
why we see a larger $\phi(M_K)$ at the bright end than seen in $K$-selected
samples.

\section{Conclusions}

We have presented preliminary estimates of the $b_J$ and $K$-band luminosity
functions obtained from CCD and infrared array imaging.
Our $b_J$ LF is consistent with an earlier estimate from photographic data.
We measure $\phi(L_K)$ over a range of 12 magnitudes, a significantly
greater range of luminosities than has been measured until now.
We are thus able to place much tighter constraints on the faint-end slope of
the $K$-band LF, albeit for a $b_J$-selected sample.
Planned future work includes measurement of the bivariate LF
$\phi(L_{b_J},L_K)$ using CCD $b_J$ magnitudes, the $R$-band LF
and luminosity functions of galaxies subdivided by colour and morphological 
type.
Given the topic of this meeting, it will also be of great interest to calculate
bivariate luminosity-surface brightness distributions.

\acknowledgements
I thank my collaborators Simon Lilly and George Efstathiou for allowing me
to show results in advance of publication, the CTIO staff for their excellent
support and Jon Gardner and Gyula Szokoly for sending me their $K$-band LF
data points.

\end{document}